**Non-stoichiometry effects on the extreme magnetoresistance in Weyl semimetal WTe$_2$**

J. X. Gong,[1,2] J. Yang,[1] M. Ge,[2] Y. J. Wang,[1] D. D. Liang,[1] L. Luo,[1] X. Yan,[1] W. L. Zhen,[1] S. R. Weng,[1] L. Pi,[1,2] C. J. Zhang,[1,3] and W. K. Zhu[1]

[1]*High Magnetic Field Laboratory, Chinese Academy of Sciences, Hefei 230031, China*

[2]*Hefei National Laboratory for Physical Sciences at Microscale, University of Science and Technology of China, Hefei 230026, China*

[3]*Institute of Physical Science and Information Technology, Anhui University, Hefei 230601, China*

Non-stoichiometry effect on the extreme magnetoresistance is systematically investigated for the Weyl semimetal WTe$_2$. Magnetoresistance and Hall resistivity are measured for the as-grown samples with a slight difference in Te vacancies and the annealed samples with increased Te vacancies. The fittings to a two-carrier model show that the magnetoresistance is strongly dependent on the residual resistivity ratio (i.e., the degree of non-stoichiometry), which is eventually understood in terms of electron doping which not only breaks the balance between electron-type and hole-type carrier densities but also reduces the average carrier mobility. Thus, compensation effect and ultrahigh mobility are probably the main driving force of the extreme magnetoresistance in WTe$_2$.

**I. INTRODUCTION**

Giant or colossal magnetoresistance (MR) is one of the most important properties in magnetic multilayers and manganites,[1,2] in which the resistance changes significantly with applied magnetic field. Such a novel property can be utilized to make multifunctional devices, e.g., magnetic sensors,[3] magnetic memory,[4] hard drives,[5] etc. While the materials with a colossal MR are mostly insulators or semiconductors, the MR in conventional metals is usually very small, showing quadratic field dependence in low fields and saturated in high fields.[6] The extremely large and non-saturated positive MR (XMR) has been recently discovered in a number of metals or semimetals, including Cd$_3$As$_2$,[7] TaP,[8] WTe$_2$,[9] PtSn$_4$,[10] antimonides,[11-16] etc.[17,18] Particularly, some of the XMR materials have been demonstrated or predicted to possess a topological semimetal state. For example, Cd$_3$As$_2$ is a three-dimensional Dirac semimetal,[19] and WTe$_2$ is a type-II Weyl semimetal.[20]

Different from the colossal MR effect, in which the carrier scattering from magnetic disorder and spin fluctuation is greatly suppressed under magnetic field,[2] various mechanisms are proposed to explain the XMR phenomena. In LaBi and LaSb, the XMR is attributed to the electron-hole compensation,[21] which means a perfect balance between the electron and hole populations. For $Cd_3As_2$ and TaAs family, the large MR cannot be simply understood in terms of the compensation effect, and one proposed scenario is that the novel topological protection suppresses the backscattering at zero magnetic field.[7,8] Spectroscopic evidence shows that the XMR in YSb is special, because YSb lacks both topological protection and electron-hole compensation.[22] The large MR of $PtSn_4$ is associated with its ultrahigh mobility.[10]

Among these systems, $WTe_2$ is a unique one whose XMR shows no sign of saturation in the magnetic field even as high as 60 T.[9] This discovery has triggered extensive researches to reveal its origin. Based on electronic structure calculation, the XMR in bulk $WTe_2$ is ascribed to the compensation between the electron and hole populations, which is supported by the angle-resolved photoemission spectroscopy (ARPES) results and the quantum oscillation experiments.[23,24] Even though, other possible origins are proposed. A distinct reduction in electron density occurs below 50 K indicating a possible electronic structure change, which might be the direct driving force of the electron-hole compensation and the XMR as well.[25] The circular dichroism observed in another ARPES experiment suggests that the spin-orbit coupling and related spin and orbital angular momentum textures play an important role in the XMR.[26] Also, the charge compensation is highly dependent on sample thickness. The balance of electron and hole states is respected only when considering at least three Te-W-Te layers,[27] which may suggest other mechanisms in thin samples. Recently, a few experiments have attempted to find out the origin by gate-tuning the carrier density of a thin sample in situ.[28-30] Until now, the origin of the XMR in $WTe_2$ is still under debate.

In this work, we try to study the non-stoichiometry effect in $WTe_2$ by comparing the samples with different Te vacancies, which are introduced during the natural growth or annealing process. MR and Hall resistivity are measured for the as-grown samples with a slight difference in Te vacancies. The fittings to a two-carrier model show that the MR is strongly dependent on the residual resistivity ratio (RRR), which is eventually understood in terms of electron doping which not only breaks the balance between electron-type and hole-type carrier densities but also reduces the average carrier mobility. Such a combined mechanism is further confirmed on the annealed samples with increased Te vacancies.

Thus, compensation effect and ultrahigh mobility are probably the main driving force of the XMR in $WTe_2$.

## II. EXPERIMENTAL METHODS

Single crystals of $WTe_2$ were grown with Te flux. In order to obtain samples of different qualities, source powders of tungsten and tellurium were prepared at a series of chemical ratios. The powder mixture was placed in an alumina crucible, with another crucible containing quartz wool mounted on top of it. Both crucibles were sealed in a silica ampoule, heated to 1000 °C and held for 10 hours, and then cooled down to 600 °C at a rate of 2 °C/h. The excess Te flux was removed by centrifugation at 600 °C. The crystal structure and phase purity were checked by single crystal X-ray diffraction (XRD) on a Rigaku-TTR3 X-ray diffractometer using Cu K$\alpha$ radiation, and the morphological characterization was performed on a Hitachi TM3000 scanning electron microscope (SEM). The chemical components of as-grown samples and annealed samples were obtained on an Oxford energy dispersive spectroscope (EDS). All the annealed samples were cut from the same piece of as-grown sample, and then annealed in vacuum for different time lengths. The Te deficiency of the as-grown sample (i.e., not annealed) was set as 0%. The deficiencies of other annealed samples were compared with the as-grown sample. The Te vacancies were obtained as 0%, 6.7% and 9.7%, respectively. The MR and Hall measurements were taken on a home-built Multi Measurement System on a Janis-9T magnet.

## III. RESULTS AND DISCUSSION

As seen in the SEM image in Fig. 1(a), the as-grown $WTe_2$ single crystal exhibits ribbon-like morphology with clean and shining surface. The single crystal XRD pattern shows only (0 0 2$l$) peaks [Fig. 1(b)], confirming that the naturally cleaved surface is the *ab* plane and the sample is a single phase, i.e., *Pmn*$2_1$ (31). Figure 1(c) illustrates the crystal structure, in which silver and orange spheres represent W and Te atoms, respectively. From Fig. 1(c), we find that $WTe_2$ adopts a typical crystal structure of layered transition metal dichalcogenides (TMDs), namely, metal layer located between adjacent chalcogenide layers. Such sandwich sheets stack along the *c* axis, with van der Waals bonding between them. As a result of the strong anisotropy, $WTe_2$ is typically electronically two-dimensional. Moreover, another structural feature is present in $WTe_2$. The zigzag tungsten chains are formed along the *a* axis, making it structurally quasi-one-dimensional.

Figure 2(a) shows the temperature dependence of resistivity for the as-grown samples of different qualities, denoted by #1, #2, #3 and #4, respectively. As described above, these samples are grown from source powders at various ratios, although the EDS characterizations show that they all keep good chemical stoichiometry without any detectable difference. The sample quality can be roughly evaluated by the residual resistivity ratio (RRR), defined as the ratio of resistivity at room temperature and at 0 K, which is strongly depending on the amounts of impurities and defects. In the present paper, the RRR takes the value $\frac{\rho(250\ K)}{\rho(2\ K)}$. As shown in Fig. 2(a), all the samples exhibit a metallic behavior with a small resistivity. However, the RRR ranges from 113 to 272, reflecting different qualities of samples, i.e., the degree of non-stoichiometry.

The longitudinal MR ($\frac{\Delta\rho}{\rho_0} = \frac{\rho_H - \rho_0}{\rho_0}$) and Hall resistivity are further measured at 2 K under a magnetic field up to 9 T, with the current applied along the *a* axis and the magnetic field applied perpendicular to the *ab* plane. The extremely large MR is reproduced on our samples, e.g., about $1.6 \times 10^5$% at 9 T for sample #1. The XMR and Hall resistivity vary significantly for different samples. Namely, with the increase of RRR, the XMR becomes larger and larger while the Hall resistivity (negative) becomes smaller and smaller. Such a positive correlation between RRR and XMR is consistent with previous report.[31] If we note that the RRR reflects the sample quality and degree of non-stoichiometry, we can deduce that the larger XMR is corresponding to the better stoichiometry. In order to check the compensation effect on XMR, a two-carrier model (or two-band model) is used to fit the MR and Hall resistivity data and extract the density and mobility of carriers. In this model, the longitudinal resistivity $\rho_{xx}$ and transverse resistivity $\rho_{xy}$ are described as[25]

$$\rho_{xx} = \frac{(n_e\mu_e + n_h\mu_h) + (n_e\mu_e\mu_h^2 + n_h\mu_e^2\mu_h)B^2}{e[(n_e\mu_e + n_h\mu_h)^2 + (n_h - n_e)^2\mu_e^2\mu_h^2 B^2]}, \tag{1}$$

and

$$\rho_{xy} = \frac{(n_h\mu_h^2 - n_e\mu_e^2)B + \mu_e^2\mu_h^2(n_h - n_e)B^3}{e[(n_e\mu_e + n_h\mu_h)^2 + (n_h - n_e)^2\mu_e^2\mu_h^2 B^2]}, \tag{2}$$

where *e* is the electron charge, *n* is the carrier density, $\mu$ is the mobility, and the subscripts e and h refer to electron-type and hole-type carriers, respectively. According to the model, the condition for $\rho_{xx}$ to increase as $B^2$ is $n_e = n_h$, which means that electrons and holes are perfectly compensated by each other. Moreover, this model also suggests that $\rho_{xy}$ should be proportional to $B$, i.e., linearly field dependent, for the perfect compensation. Slight deviation from the perfect compensation will cause a nonlinear field dependence of Hall resistivity in high field region. A simple power-law fitting for sample #1 (which has the best quality) gives MR$\propto B^{1.77}$ [as shown in Fig. 2 (b)], as well as the nonlinear $\rho_{xy}(B)$ curves, suggesting that the samples are not in the perfectly compensated regime. Further fittings using the above equations result in the carrier densities ($n_e$, $n_h$) and carrier mobilities ($\mu_e$, $\mu_h$). Non-oscillatory curves in Fig. 2(c) represent the fitted curves, which agree well with the experimental data, showing sufficiently good fittings and reliable fitting results.

Figure 3(a) presents the MR and $n_e/n_h$ as a function of RRR for all the as-grown samples, obtained from Fig. 2. It is easily found that with the increase in RRR and XMR, the ratio $n_e/n_h$ gradually approaches the perfect compensation value, i.e., 1. That is to say, the dependence of XMR on the RRR can be indeed interpreted as the change of $n_e/n_h$. For a sample of higher quality, the less defects correspond to a larger RRR, and the more balanced $n_e/n_h$ ratio gives rise to a larger XMR. Similar phenomenon has been found in the investigation on the temperature dependence of XMR.[25]

Since the XMR is strongly dependent on the non-stoichiometry effect, artificial defects may be introduced to conduct more conclusive research work, considering that the naturally as-grown samples are very close to each other in stoichiometry but a distinct gradient of vacancies can be obtained by the annealing processes. Different time lengths of annealing in vacuum yield a series of non-stoichiometric samples, i.e., 0%, 6.7% and 9.7%. The percentages represent the degree of Te vacancies, setting the as-grown sample as 0%. Note that the three samples are cut from the same piece of as-grown sample. As shown in Figs. 4(a)-4(c), with the increase of Te vacancies, the RRR and XMR are greatly reduced. The degradation of sample quality indeed suppresses the XMR. In order to reveal the underlying information, again we fit the MR and Hall resistivity data with the two-carrier model. The resultant $n_e/n_h$ ratios are plotted as a function of Te vacancies in Fig. 3(b), along with the XMR. We can see that the introduced defects in the annealed samples indeed lead to a less compensated $n_e/n_h$, which is eventually responsible for the reduced XMR. The results obtained from the annealed samples are consistent with those from the as-grown samples.

Note that the information of carrier concentrations can also be extracted from the quantum oscillations revealed in the MR and Hall resistivity. The Hall resistivity is obtained by averaging the data taken under positive and negative fields, to remove the possible mixing of longitudinal MR (also including the oscillations in MR) due to the nonsymmetrical Hall contacts. In view of this, the Hall resistivity may be not as suitable as the MR data for the quantum oscillation analyses. So we analyze the Shubnikov-de Haas (SdH) oscillations in the MR to give an estimate of the carrier concentrations. After subtracting the non-oscillatory background from the MR data and employing fast Fourier transformation (FFT), four major frequencies (i.e., 90 T, 123 T, 140 T and 160 T) can be found in all of the samples, which are consistent with other research work on WTe$_2$.[24] However, no systematic difference that is beyond reasonable doubt can be detected among these samples, as shown in Figs. 5(a) and 5(b).

For simplicity, a rough estimate can be given based on the free electron assumption, i.e., spherical Fermi surfaces for all the carrier pockets. The Fermi radii and carrier concentrations of four pockets are calculated and summarized in Table I. We note that these values are generally consistent with those obtained from the two-carrier model fits. They have the same order of magnitude, i.e., $10^{19}$ cm$^{-3}$. Moreover, the total electron-type carrier concentration ($1.7 \times 10^{19}$ cm$^{-3}$) is roughly compensated by the hole-type one ($1.62 \times 10^{19}$ cm$^{-3}$). The slightly larger electron-type carrier concentration agrees with the situation of electron doping, due to the Te deficiency. All these facts suggest that our measurements and analyses are convincing.

Here we propose a possible scenario to explain the effect of carrier concentrations on the XMR in WTe$_2$. From the band structure calculations and quantum oscillation analyses,[23,24] the Fermi surfaces can be generally described by the two-carrier model, i.e., one electron-like band and one hole-like band. As illustrated in Fig. 6(a), the band structures around the Fermi level are composed of one electron-type pocket and one hole-type pocket. When the electron-type carriers and hole-type carriers are completely compensated, which is also the stoichiometric case, the MR reaches its maximum value, according to the above equations. However, once the electron-type carriers and hole-type carriers become non-compensated, no matter what reason it is, the MR is reduced. In the present experiment, both the as-grown samples and the annealed samples have a certain degree of Te vacancies, i.e., electron doping. As seen in Fig. 6(b), the electron doping increases the Fermi level, which breaks the balance between electron-type carriers and hole-type carriers and leads to a reduction in MR. A further reasonable guess is that the opposite situation, i.e., the hole doping with the hole-type carriers exceeding the electron-type carriers, may also comply with this rule.

In addition to the non-compensation of carrier densities, the non-stoichiometry effect may also lead to the change of carrier mobilities, because the existence of impurities and disorders would increase the scattering. The average carrier mobility ($\mu_{ave}$) is found to strongly correlate with sample quality (i.e., RRR) and MR ratio.[31] However, the analysis in that paper is based on strongly simplified assumptions, i.e., $n_h = n_e = n$ and $\mu_h \approx \mu_e \approx \mu$, which are however impossible in actual cases. More seriously, these assumptions naturally lead to the relation of $\rho_{xx} \propto 1/\mu + \mu B^2$, and further MR ratio $= 1 + (\mu B)^2$ and RRR $= \rho_{300K}/\rho_{2K} \propto \mu_{2K}$. The authors use this relation to fit the MR data and obtain the average mobility $\mu_{ave}$, and deduce the linear correlation of $\mu_{ave}$ with RRR. Here we may find that the result has been already included in the initial assumption, which is inappropriate during data analysis. To solve this problem, we adopt the values of carrier mobilities that are obtained from the two-carrier model fittings, and calculate the average mobility $\mu_{ave}$ with different weights (i.e., carrier densities $n$) of holes and electrons. As shown in Fig. 3(c), with the increase of RRR, the $\mu_{ave}$ indeed increases, consistent with Ref. [31]. But we are not sure if the correlation is in a linear way, because of the limited data points. Such a correlation is confirmed on the annealed samples; namely, the $\mu_{ave}$ is suppressed by the increase of Te vacancies [Fig. 3(d)].

These results suggest that the existence of impurities or disorders causes not only the non-compensation of carrier densities but also the reduction of average carrier mobility, both of which play an important role in the XMR. The largest MR is arising from the combined effect of the most compensated carriers and the highest average mobility.

Before ending the present paper, we may talk more about the relationship between XMR and topological semimetal states. Although all the found topological semimetals seem to exhibit the XMR effect, the XMR does not necessarily lead to topological semimetal states. For some XMR materials like LaSb,[12] distinct trivial features are revealed by first-principles calculations and ARPES experiment.[21,32] In such systems, the XMR may be just originating from the compensation effect and/or ultrahigh carrier mobilities, not relevant to any topological mechanism.

## IV. CONCLUSIONS

In conclusion, the non-stoichiometry effect on the XMR is investigated for the Weyl semimetal

WTe$_2$. MR and Hall resistivity are measured for the as-grown samples with a slight difference in Te vacancies and the annealed samples with increased Te vacancies. The fittings to a two-carrier model show that the MR is strongly dependent on the RRR (i.e., the degree of non-stoichiometry), which is eventually understood in terms of electron doping which not only breaks the balance between $n_e$ and $n_h$ but also reduces the average mobility $\mu_{ave}$. Thus, compensation effect and ultrahigh mobility are probably the main driving force of the XMR in WTe$_2$.

**ACKNOWLEDGMENTS**

This work was supported by the National Key R&D Program of China (Grant Nos. 2016YFA0300404 and 2017YFA0403600), the National Natural Science Foundation of China (Grant Nos. 51603207, U1532267, 11574288 and 11674327), and the Natural Science Foundation of Anhui Province (Grant No. 1708085MA08).

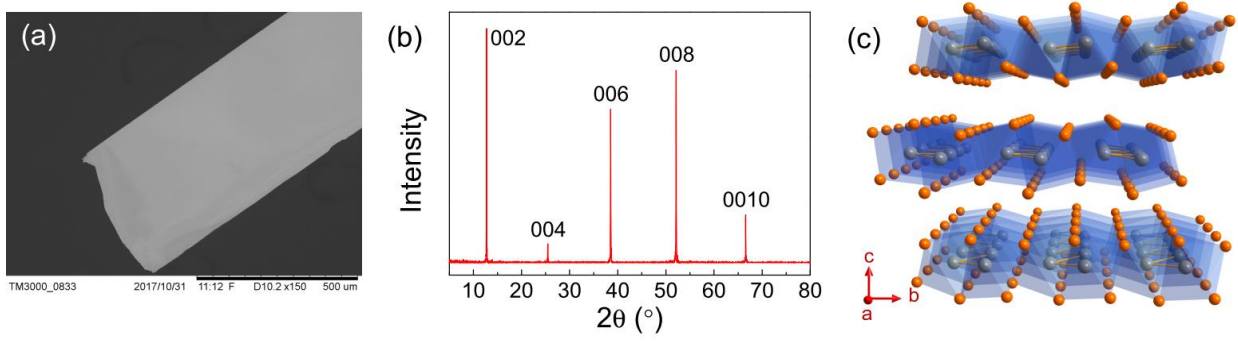

FIG. 1. (a) SEM image of WTe$_2$ single crystal. (b) Single crystal XRD pattern. (c) Crystal structure of WTe$_2$ along the *a* axis that is parallel to the W-W zigzag chains. Silver and orange spheres represent W and Te atoms, respectively.

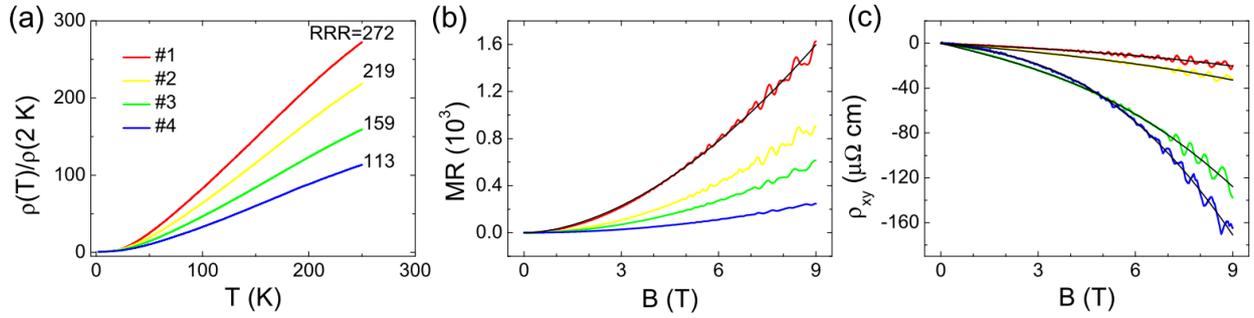

FIG. 2. (a) Temperature dependence of resistivity normalized by $\rho(2\text{ K})$ for samples #1, #2, #3 and #4, with the current applied along the *a* axis. Residual resistivity ratio (RRR), defined as $\rho(250\text{ K})/\rho(2\text{ K})$, is also presented. (b) Longitudinal magnetoresistance and (c) transverse Hall resistivity taken at 2 K in a field range of 0-9 T. Non-oscillatory curve in (b) represents the power-law fitting; non-oscillatory curves in (c) represent the fittings with the two-carrier model.

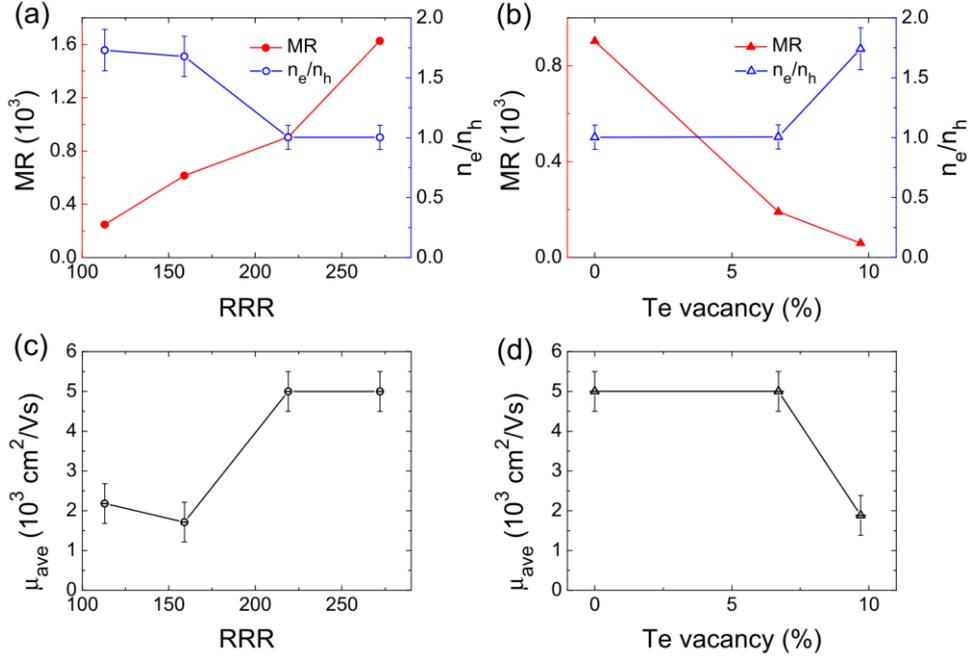

FIG. 3. (a) MR measured at 2 K and 9 T (filled circles), the fitting-resultant $n_e/n_h$ (open circles), and (c) average mobility $\mu_{ave}$ (strikethrough circles) as a function of RRR for the as-grown samples, obtained from Fig. 2. (b) MR (filled triangles), $n_e/n_h$ (open triangles), and (d) $\mu_{ave}$ (strikethrough triangles) as a function of Te vacancy for the annealed samples, obtained from Fig. 4.

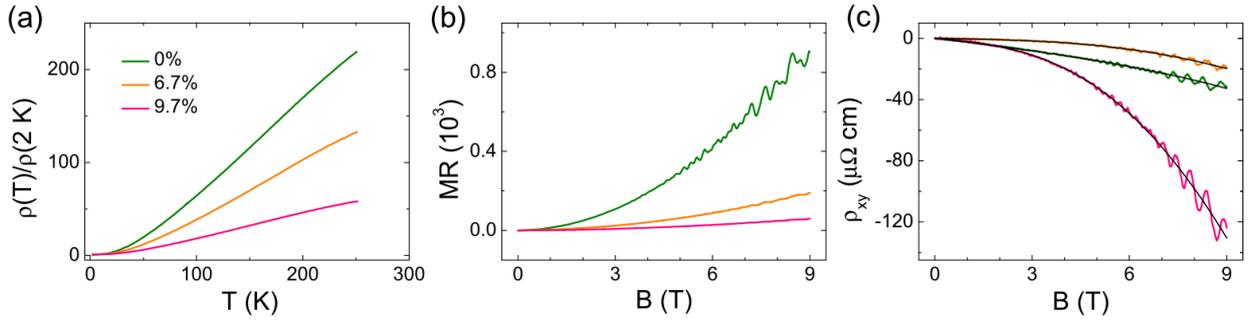

FIG. 4. (a) Temperature dependence of resistivity normalized by $\rho(2\ K)$ for the annealed samples with different Te vacancies. (b) Longitudinal magnetoresistance and (c) transverse Hall resistivity taken at 2 K in a field range of 0-9 T. Non-oscillatory curves represent the fittings with the two-carrier model.

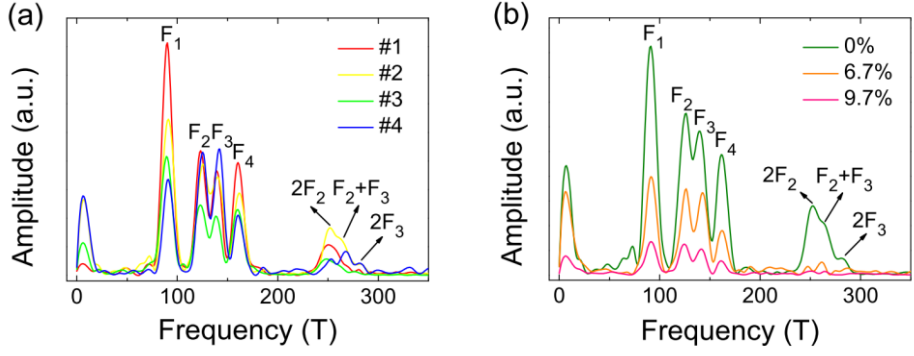

FIG. 5. FFT spectra of the SdH oscillations for (a) as-grown samples and (b) annealed samples.

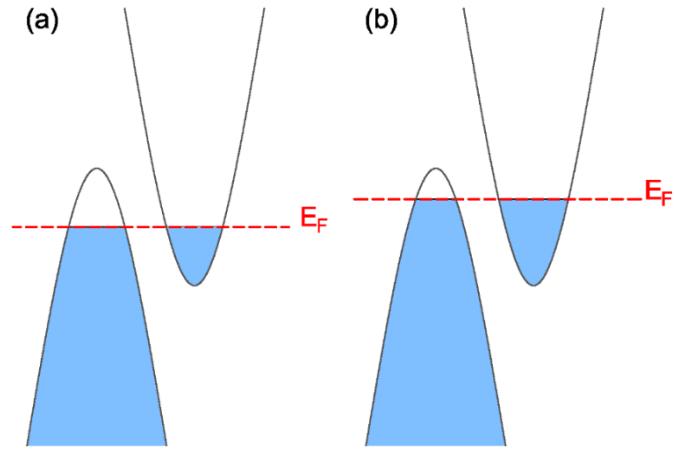

FIG. 6. Schematic illustration for the non-compensation effect in WTe$_2$. Dashed lines indicate the Fermi level. Condition in (a) means perfect compensation of electron-type and hole-type carriers. (b) With electron doping, the electron-type carriers exceed the hole-type carriers. The less compensated ratio of $n_e/n_h$ leads to a decrease in the XMR of WTe$_2$.

TABLE I. Quantum oscillation frequencies, the calculated Fermi radii and carrier concentrations of four pockets in Fig. 5. The symbols *h* and *e* refer to hole-type and electron-type pockets, respectively.

|  | $F$ (T) | $k_F$ (nm$^{-1}$) | $n$ ($10^{19}$ cm$^{-3}$) |
| --- | --- | --- | --- |
| $F_1$ (*h*) | 90 | 0.52 | 0.48 |
| $F_2$ (*e*) | 123 | 0.61 | 0.77 |
| $F_3$ (*e*) | 140 | 0.65 | 0.93 |
| $F_4$ (*h*) | 160 | 0.70 | 1.14 |